\documentclass[12pt]{article}
\usepackage{graphicx}
\usepackage{url}

\newcommand\pubdate{\today}

\newcommand\pubnumber{change/delete REPORT-\#}

\def\Title#1{\begin{center} {\Large #1 } \end{center}}
\def\Author#1{\begin{center}{ \sc #1} \end{center}}
\def\Address#1{\begin{center}{ \it #1} \end{center}}

\newcommand\pubblock{\rightline{\begin{tabular}{l} \pubnumber\\
         \pubdate  \end{tabular}}}
\newenvironment{Abstract}{\begin{center}{\bf Abstract}\end{center} \bigskip \begin{quotation}  }{\end{quotation}}
\newenvironment{Presented}{\begin{quotation} \begin{center} 
             PRESENTED AT\end{center}\bigskip 
      \begin{center}\begin{large}}{\end{large}\end{center} \end{quotation}}

\def\beq{\begin{equation}}
\def\eeq#1{\label{#1}\end{equation}}
\def\eeqn{\end{equation}}

\def\beqa{\begin{eqnarray}}
\def\eeqa#1{\label{#1}\end{eqnarray}}
\def\eeqan{\end{eqnarray}}

\let\bar=\overbar

\def\Dslash{\not{\hbox{\kern-4pt $D$}}}
\def\dslash{\not{\hbox{\kern-2pt $\del$}}}

\def\BR{\mbox{\rm BR}}

\def\msb{{\bar{\ssstyle M \kern -1pt S}}}

\newcommand {\bbar}	{\ensuremath{\bar b}}
\newcommand {\bbbar}	{\ensuremath{b\bbar}}
\newcommand {\pbar}	{\ensuremath{\bar p}}
\newcommand {\ppbar}	{\ensuremath{p\pbar}}
\newcommand {\tbar}     {{\ensuremath{\bar t}}}
\newcommand {\ttbar}    {{\ensuremath{t\tbar}}}
\newcommand {\GeV}        {{\ensuremath{\,\mathrm {GeV}}}}

\newcommand {\ifb}        {{\ensuremath{\,\mathrm {fb}^{-1}}}}
\newcommand {\mttbar} {{\ensuremath{m_\ttbar}}}
\newcommand {\emiss}{/\!\!\!\!E} 
\newcommand {\met} {\ensuremath{{\emiss_{\mathrm T}}}}
\newcommand {\pt} {\ensuremath{{p_{\mathrm T}}}}

\newcommand {\cts} {\ensuremath{{\cos\theta^*}}}
\newcommand {\mzp} {\ensuremath{{M_{Z'}}}}
\newcommand {\thw} {\ensuremath{{\theta_W}}}
\newcommand {\mwp} {\ensuremath{{M_{W'}}}}
\newcommand {\mnr} {\ensuremath{{m(\nu_R)}}}

\newcommand {\pythia}   {{\sc pythia}}
\newcommand {\madgraph}   {{\sc madgraph}}

\begin{document}
\begin{titlepage}
\pubblock

\vfill


\Title{New Physics in Top at Tevatron}
\vfill
\Author{A. Harel, for the CDF and D0 Collaborations}  
\Address{University of Rochester, Rochester, NY, 14627-0171, USA}
\vfill


\begin{Abstract}
I report on direct and indirect searches for new physics in top
events from the CDF and D0 Collaborations at the Fermilab Tevatron
Collider.
\end{Abstract}

\vfill

\begin{Presented}
The Ninth International Conference on\\
Flavor Physics and CP Violation\\
(FPCP 2011)\\
Maale Hachamisha, Israel,  May 23--27, 2011
\end{Presented}
\vfill

\end{titlepage}
\def\thefootnote{\fnsymbol{footnote}}
\setcounter{footnote}{0}
%


\section{Introduction}

The top quark is the heaviest known elementary particle,
making it a natural probe for new physics. In  addition,
it is the only known elementary particle whose mass is near
the scale of electroweak symmetry breaking, suggesting that
it plays a special role there.
The top quark was discovered by the CDF and D0 Collaboration in Run I 
of the Fermilab Tevatron Collider~\cite{bib:discovery}.
In Run II, the Tevatron collides protons and antiprotons
with a center of mass energy of $1.96\ifb$,
providing large samples of top quark that enable CDF and D0
to search for signs of new physics in top quark events.
The CDF and D0 detectors are described in Ref.~\cite{bib:dets}.

I report on recent direct searches for processes beyond the 
standard model (BSM), 
and on indirect searches performed by measuring top quark 
properties which are firmly predicted in the standard model (SM),
and looking for significant deviations from these predictions.
Measurements of asymmetries in \ttbar\ production fall into
the latter category, but are reported elsewhere in these
proceedings~\cite{bib:schwarz}.

\section{\boldmath Searches for narrow \ttbar\ resonances}

Both Tevatron collaborations search for ``narrow'' resonances
in \ttbar\ production, that is, resonances in the \mttbar\ spectrum
with an intrinsic width that is smaller than the experimental resolution
for \mttbar~\cite{bib:mttres}. 
This is implemented by taking the width to be $1.2$\% of
the resonance mass.

The searches are performed in the ``lepton+jets'' channel, where
on of the two $W$ bosons from the $t\to Wb$ decays~\footnote
{charged conjugate processes are included implicitly throughout.}
decayed leptonically into $l\bar{\nu}_l$ and the other $W$ boson
decayed hadronically into $q\bar{q'}$. When the isolated lepton from 
the leptonic $W$ boson decays is an electron or muon, even if produced
from an intermediate $\tau$ lepton, it provides a strong experimental
handle. Thus hadronic $\tau$ lepton decays are not considered as part
of the signal in this channel.

Data samples are further enriched in top decays by requiring
that the events contain several jets, as typically three or four jets
arise from the four quarks produced in these decays,
an imbalance in transverse energy (\met) due to the neutrino, and
displaced decay vertices typical of the hadronization and subsequent
electroweak decay of the $b$ quarks. The resulting sample
 compositions are shown in Fig.~\ref{fig:mtt-spectra}.

\begin{figure}[htb]
\centering
\includegraphics[width=0.45\textwidth]{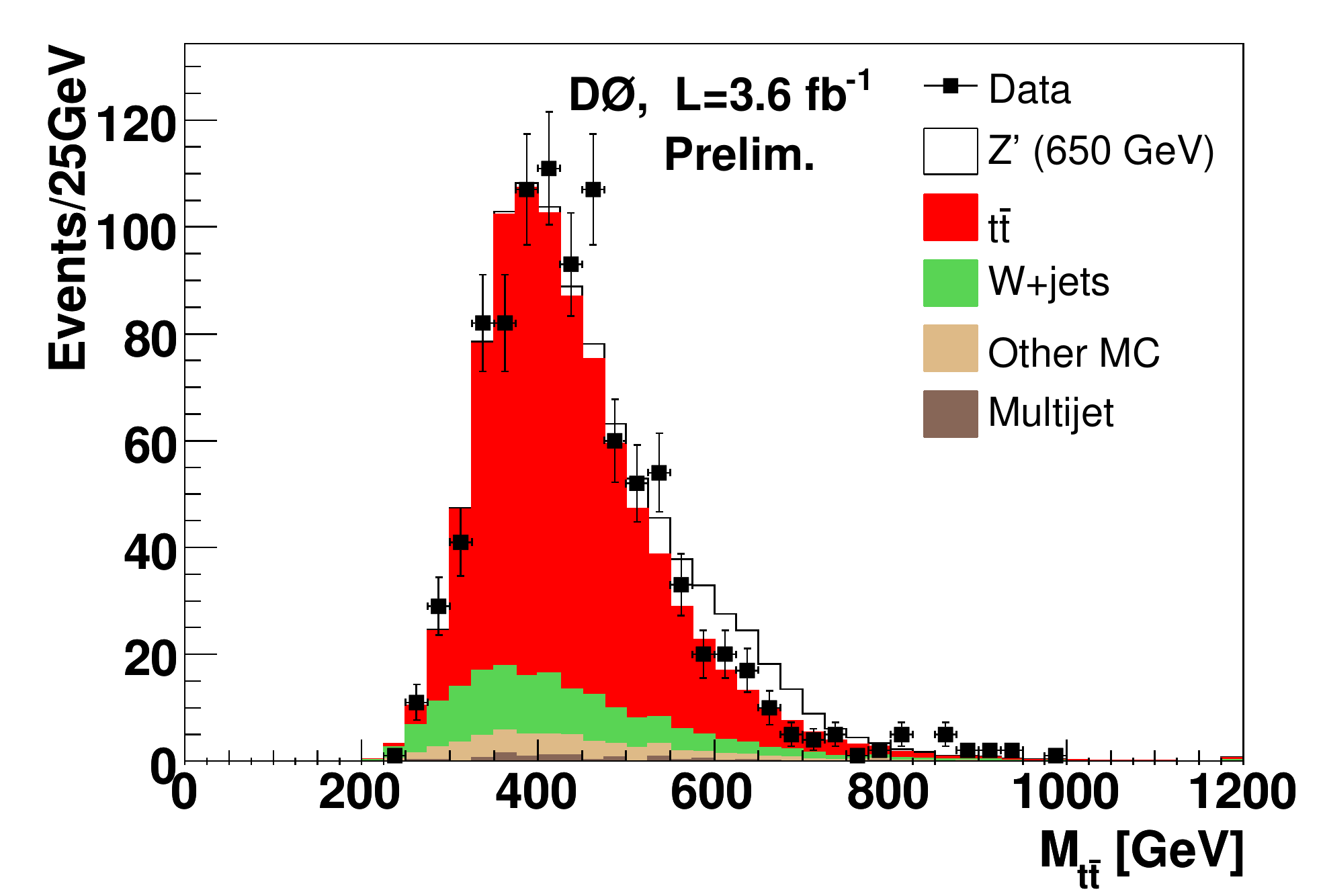}
\hspace{0.02\linewidth}
\includegraphics[width=0.51\textwidth]{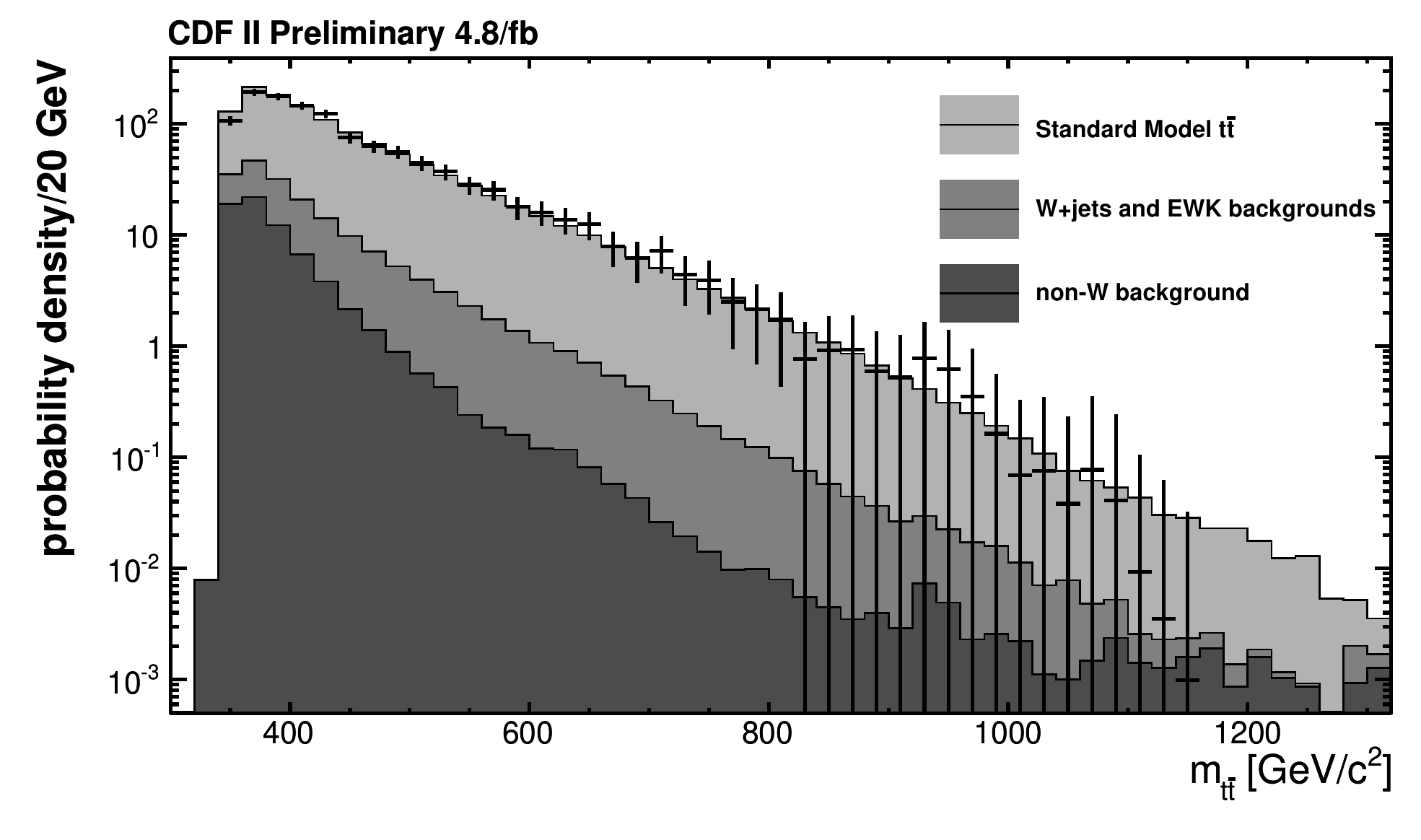}
\caption{Reconstructed \mttbar\ spectra and sample compositions
from the D0 (left) and CDF (right) searches for resonant
\ttbar\ production.}
\label{fig:mtt-spectra}
\end{figure}

The background arise mainly from $W$+jets and multijet production.
The former is enriched in heavy-flavor jets, by the requirement
that jets are tagged as arising from $b$ quarks.
It is simulated using a matched tree-level Monte Carlo (MC) generator~\cite{bib:alpgen}
coupled with the \pythia\ shower generator~\cite{bib:pythia}, and with the
heavy-flavor content increased to match data.
Other sources of background, such as single-top and di-boson
production, are modeled from simulation.

The multijet background arises when a jet is misidentified as an isolated lepton.
For example, this happens when a lepton produced in the decay of one
of the jet's hadrons carries most of the jet's energy, and the rest
of the jet is not reconstructed, so that the lepton is correctly
reconstructed, but its identification as an isolated lepton is false.
This background is modeled using auxiliary data samples.

In the D0 search, \mttbar\ is taken from 4-vector sum of the jets,
lepton, and neutrino candidate, where the neutrino is reconstructed
from the \met\ using the measured lepton 4-vector and the known $W$ boson mass.
This simple reconstruction technique works equally well when only
three jets are selected, and these events are included in the D0 search.

In the CDF search, rather than reconstruct one \mttbar\ value per event,
the probability density for each \mttbar\ value is calculated per event using
the observed 4-vectors, the SM matrix elements for such \ttbar\ production, 
the parton distribution functions (PDFs), and the transfer functions. 
The transfer functions account for the differences between the observed jets
and their progenitor quarks.
This includes both SM effects from Quantum Chromodynamics (QCD), such as
showering and hadronization, and the experimental resolution for the
detected objects. The reconstructed spectrum for each sample is
then the sum of the per-event probability densities.

To increase statistical strength, each search is performed in
several channels, by lepton type, number of jets, and number
of $b$-tagged jets.
Then the signal cross section is measured for each possible
resonance mass using a maximal likelihood estimate, and Bayesian
limits are set at the 95\% (and 68\%) CL. 
Continuum \ttbar\ production is a background to the resonant
production being measured, and is estimated using either matched MC (in D0) or 
using \pythia\ with $k$ factors to match NLO calculations (in CDF).
Systematic uncertainties
are included using nuisance parameters which are integrated out
in the limit setting.
The results are shown in Fig.~\ref{fig:mtt-res}, together with the
mass limits for a specific type of resonance.
D0 exclude $\mzp<820\GeV$, while CDF exclude $\mzp<900\GeV$.

\begin{figure}[htb]
\centering
\includegraphics[width=0.45\textwidth]{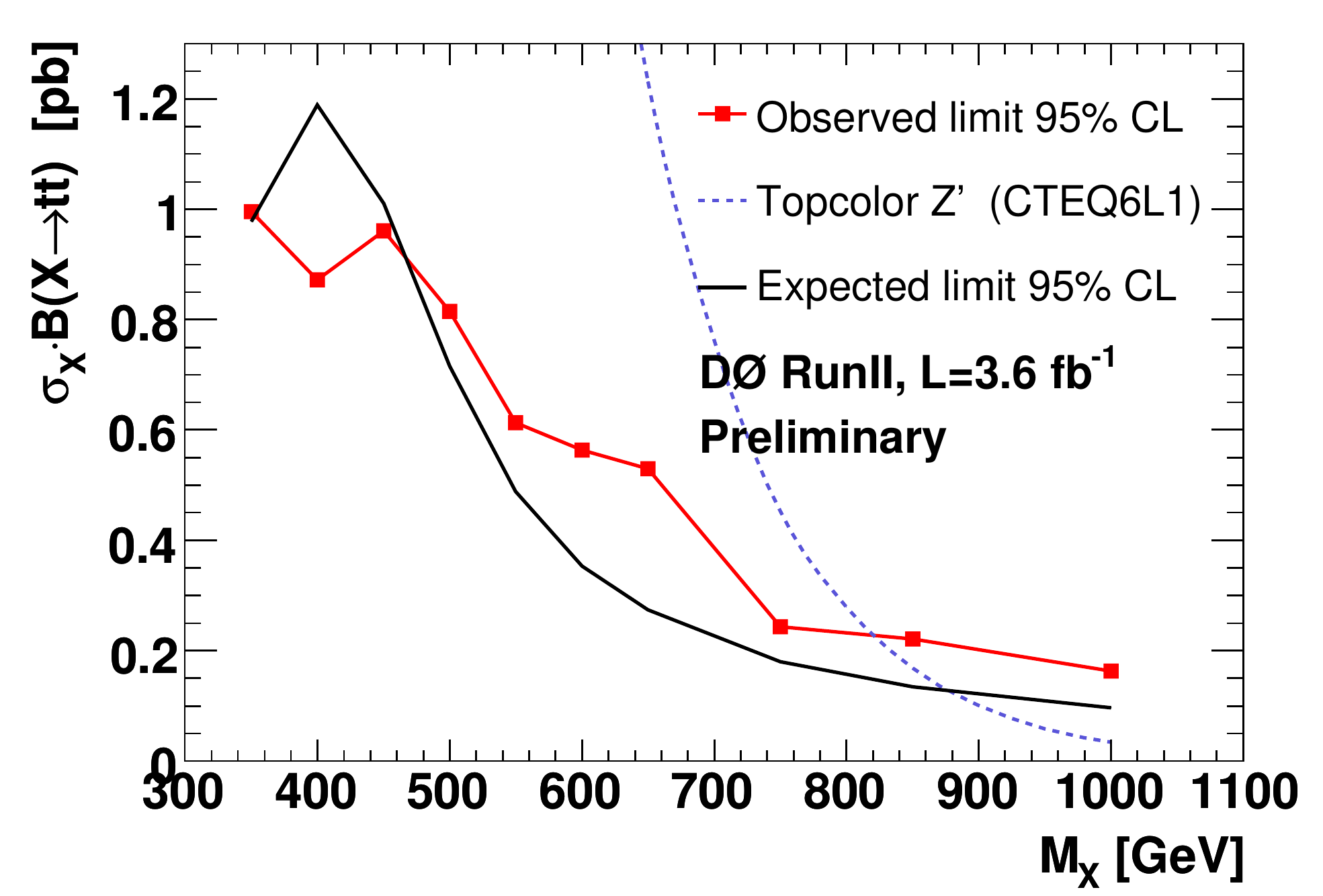}
\hspace{0.02\linewidth}
\includegraphics[width=0.51\textwidth]{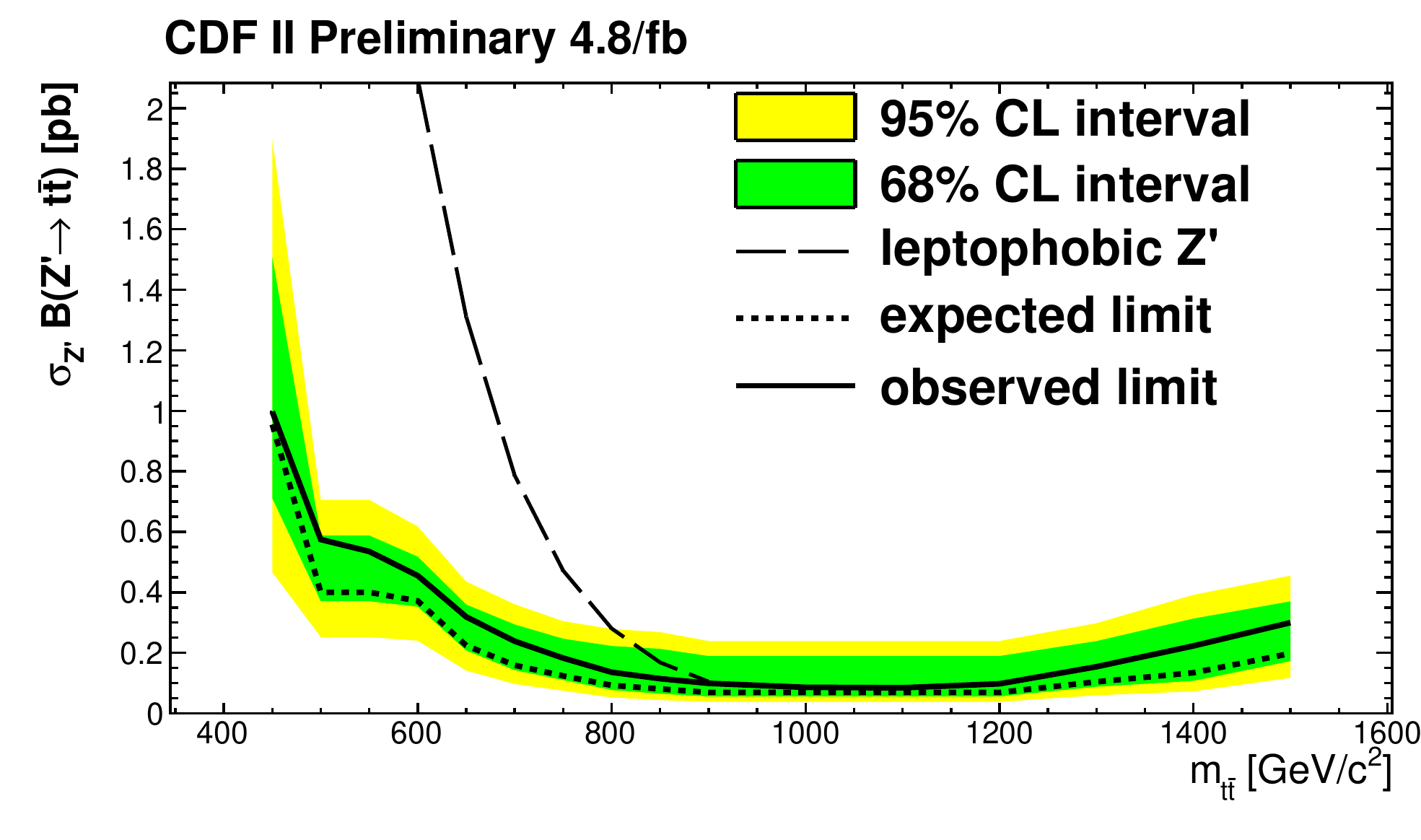}
\caption{Reconstructed \mttbar\ spectra and sample compositions
from the D0 (left) and CDF (right) searches for resonant
\ttbar\ production.}
\label{fig:mtt-res}
\end{figure}

\section{Flavor-Changing Neutral Current in top decays}

In the SM, flavor-changing neutral current (FCNC) decays of top quarks
are extremely rare: $\BR\left(t\to Zc\right)$ is of order $10^{-14}$,
and $\BR\left(t\to Zu\right)$ is of order $10^{-17}$.
FCNC decays can be enhanced by BSM processes, such as SUSY
or quark compositeness, which can reach branching fractions
of order per mille. 

In the top sector, FCNC vertices that include a gluon are
constrained by D0 using single top events~\cite{bib:singletop-FCNC}.
Those that include a $Z$ boson were previously constrained by CDF in the 
$Z$+$4\,$jets channel to have $\BR\left(t\to Zq\right)<3.7$\%,
with an expected limit of $5.0$\%~\cite{bib:CDF-FCNC}.

Recently, D0 published a search for FCNC decays that include a
$Z$ boson in the tri-lepton channel (diagrammed in Fig.~\ref{fig:fcnc-feynman}),
where both $Z$ and $W$ bosons decay leptonically~\cite{bib:DZ-FCNC}.
This final state offers excellent background rejection, but
suffers from low rates. To maximize the selection efficiency, 
electron candidates are considered out to $\left|\eta\right|=2.5$,
which requires a special reconstruction algorithm in the 
inter-cryostat region (ICR), $1.1<\left|\eta\right|<1.5$.
The background rejection of the ICR electron candidates is relatively
weak, so they are used only for the $Z\to e e$ decay, where
excellent background rejection is available using  the
known $Z$ boson mass. The muon coverage extends to $\left|\eta\right|=2.0$.
The sample composition is shown in Fig.~\ref{fig:fcnc-njet}.

\begin{figure}
\begin{minipage}[t]{0.46\linewidth}
 \centering
 \includegraphics[width=0.75\textwidth]{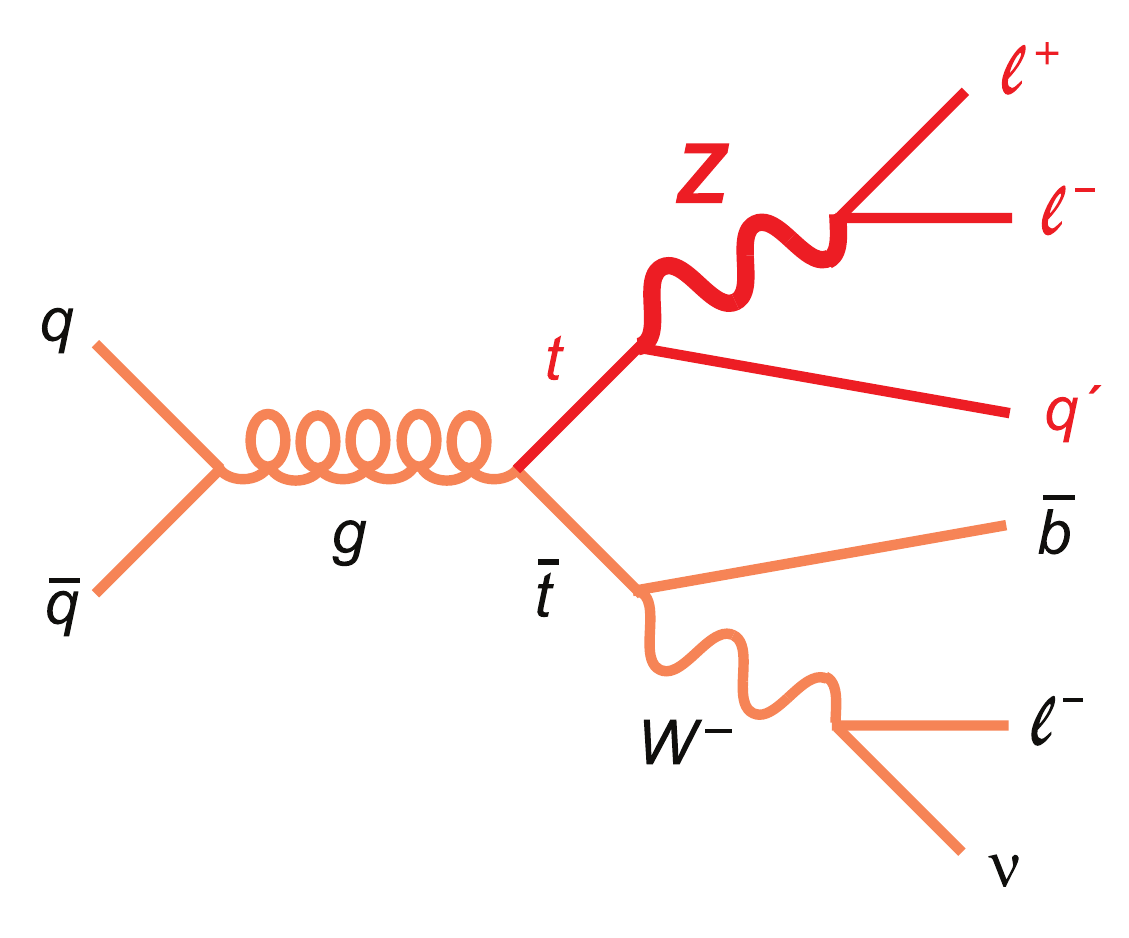}
 \caption{\label{fig:fcnc-feynman}
   Feynman diagram for the tri-lepton FCNC top decay.}
\end{minipage}
\hspace{0.02\linewidth}
\begin{minipage}[t]{0.51\linewidth}
 \centering
 \includegraphics[width=\textwidth]{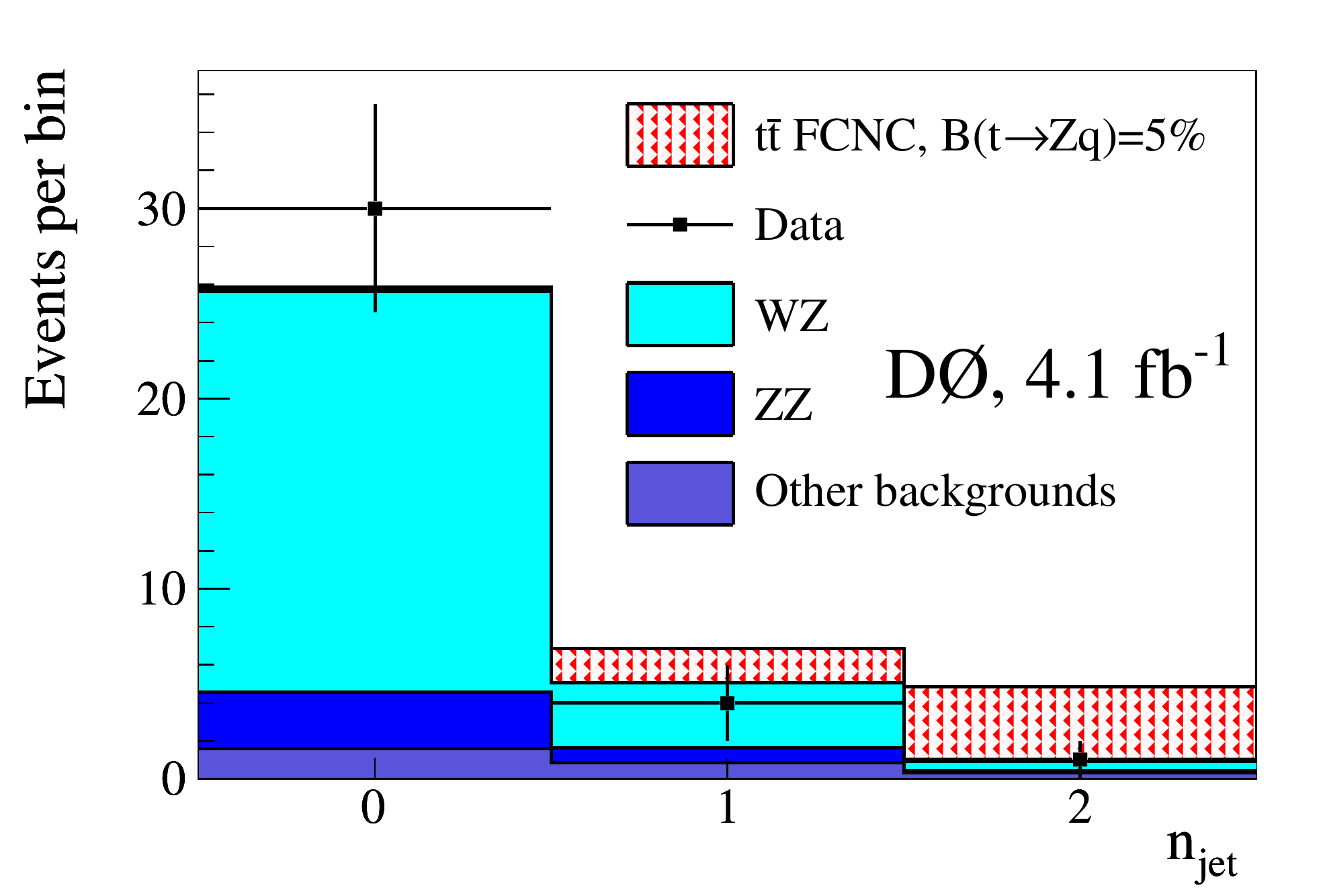}
 \caption{\label{fig:fcnc-njet}
   Number of jets and sample composition in the FCNC decay search.}
\end{minipage}
\end{figure}

D0 search for FCNC decays described by the term:
\begin{equation}
{\cal L}_{\mathrm FCNC} = \frac{e}{2 \sin \thw \cos \thw}
{\bar t}\gamma_\mu \left( v_{tqZ} - \gamma_5 a_{tqZ} \right) q Z^\mu + {\mathrm h.c.},
\end{equation}
where $e$ is the electron charge, $\thw$ is the Weinberg angle,
and $t$, $q$, and $Z$ are the fields whose quanta are the quarks and the $Z$ boson.
To test for the presence of signal, D0 uses the log of the ratio between the 
likelihood of the signal under the signal+background and the background only hypotheses.
The likelihoods are calculated for nine $H_T$ distributions, where $H_T$ is
the scalar sum of the transverse momenta (\pt) of the leptons, jets, and \met\
in each event. These distributions are for all events with no jets,
and for events with either one or more than one jets, which are divided
into four ranges in reconstructed top mass ($<120$, $120$--$150$, $150$--$200$, and $>200\GeV$).
The separation between signal and background is demonstrated in Fig.~\ref{fig:fcnc-sep}.

\begin{figure}[htbp]
  \centering
  \includegraphics[width=0.5\textwidth]{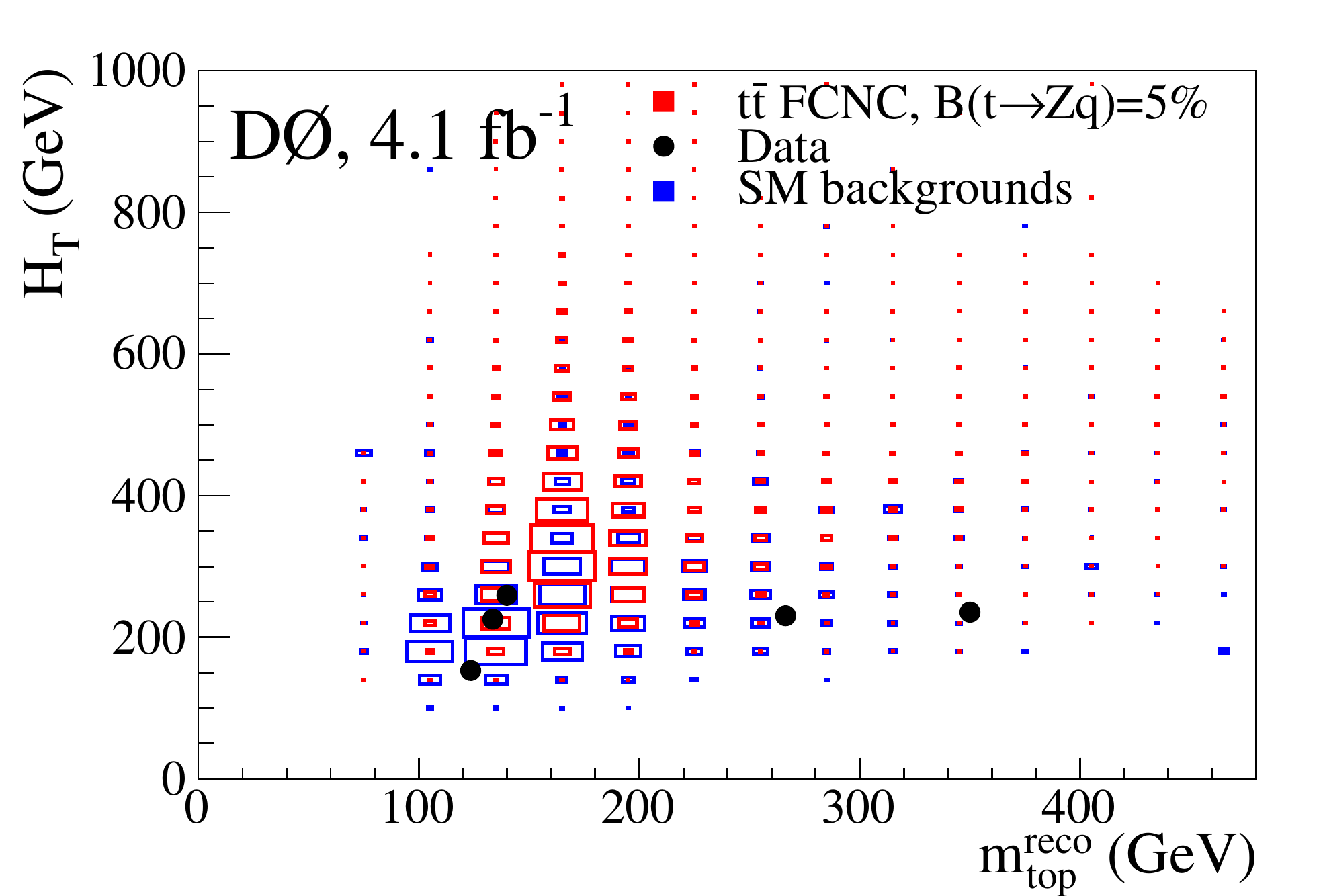}
 \caption{\label{fig:fcnc-sep}
   Discriminating variables for events with $\geq1$ jet in the FCNC decay search.}
\end{figure}

Limits are set on $\BR\left(t\to Zq\right)$ using the CLs method~\cite{bib:cls},
with systematic uncertainties described using nuisance parameters that
are constrained to data using a generalized $\chi^2$ test statistic~\cite{bib:fisher}.
The limits have only a slight dependence on the chiral structure of the resonance,
which is treated as a source of systematic uncertainty so the that the limits
apply to resonances of any chiral structure. 
D0 find a limit of $\BR<3.2$\%, with an expected limit of $3.8$\%.

\section{\boldmath $W'\to tb$ search}

The D0 Collaboration published a search for a heavy copy of the 
$W$ boson, a $W'$~\cite{bib:wtb}. The interaction between the 
$W'$ boson and the SM is given, in the most general form of lowest
dimension, by the term
\begin{equation}
{\cal L}_{W'} =
\frac{V_{ij}g_w}{2\sqrt{2}}\bar{f}_i\gamma_\mu\left[a^R_{ij}\left(1+\gamma^5\right) 
+ a^L_{ij}\left(1-\gamma^5\right)\right]W'^\mu f_j+{\mathrm{h.c.}},
\end{equation}
where $V$ is the CKM matrix for quarks and the identity matrix for leptons,
$g_W$ is the SM weak coupling constant, and $a^L$ and $a^R$ are the
left- and right-handed couplings of the $W'$ field (whose quanta is the $W'$ boson)
and the fermion doublet fields, $f$.
The left-handed coupling are motivated by the possibility of
Kaluza-Klein (KK) excitation of a $W$ boson, which would have the
same couplings as the $W$ boson of the SM.
The right-handed couplings are motivated by left-right symmetric models,
which imply that the $W'$ couples dominantly to right-handed fermions.

The search is for 
the process $\ppbar\to W'X; W'\to t b; t\to Wb; W\to l\nu$.
This process is similar to the SM s-channel single top production but
the intermediate boson is a $W'$ rather than an off-shell $W$.
Thus this search builds upon D0's single top observation~\cite{bib:D0singletop},
using a dataset corresponding to $2.3\ifb$ of integrated luminosity.
The same 24 channels are used, with the data divided by lepton flavor,
number of jets, number of $b$-tags, and data taking period.
As in Ref.~\cite{bib:D0singletop}, single top production is identified
using boosted decision tree discriminants. The discriminants combine 49 input variables, and
were trained using $a^L=a^R=1$ for each channel and for each $W'$ mass considered.
The resulting discrimination and the sample composition are shown in Fig.~\ref{fig:wp-sep}.

\begin{figure} [htbp]
\begin{minipage}[t]{0.48\linewidth}
 \centering
 \includegraphics[width=\textwidth]{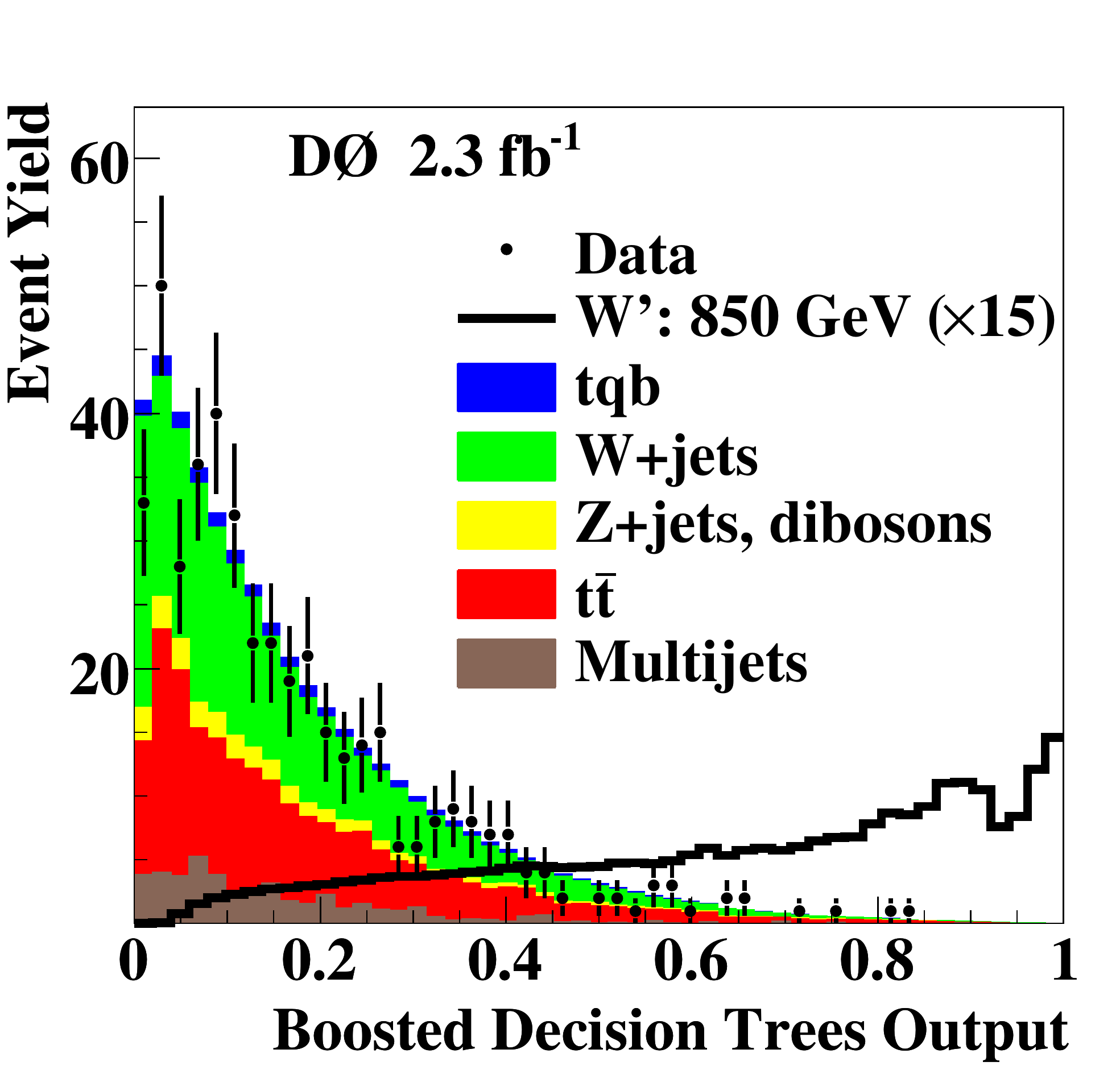}
 \caption{\label{fig:wp-sep}
   Discriminant for $W'$ search and sample composition.}
\end{minipage}
\hspace{0.03\linewidth}
\begin{minipage}[t]{0.48\linewidth}
 \centering
 \includegraphics*[width=0.99\textwidth,viewport=20 00 580 530]{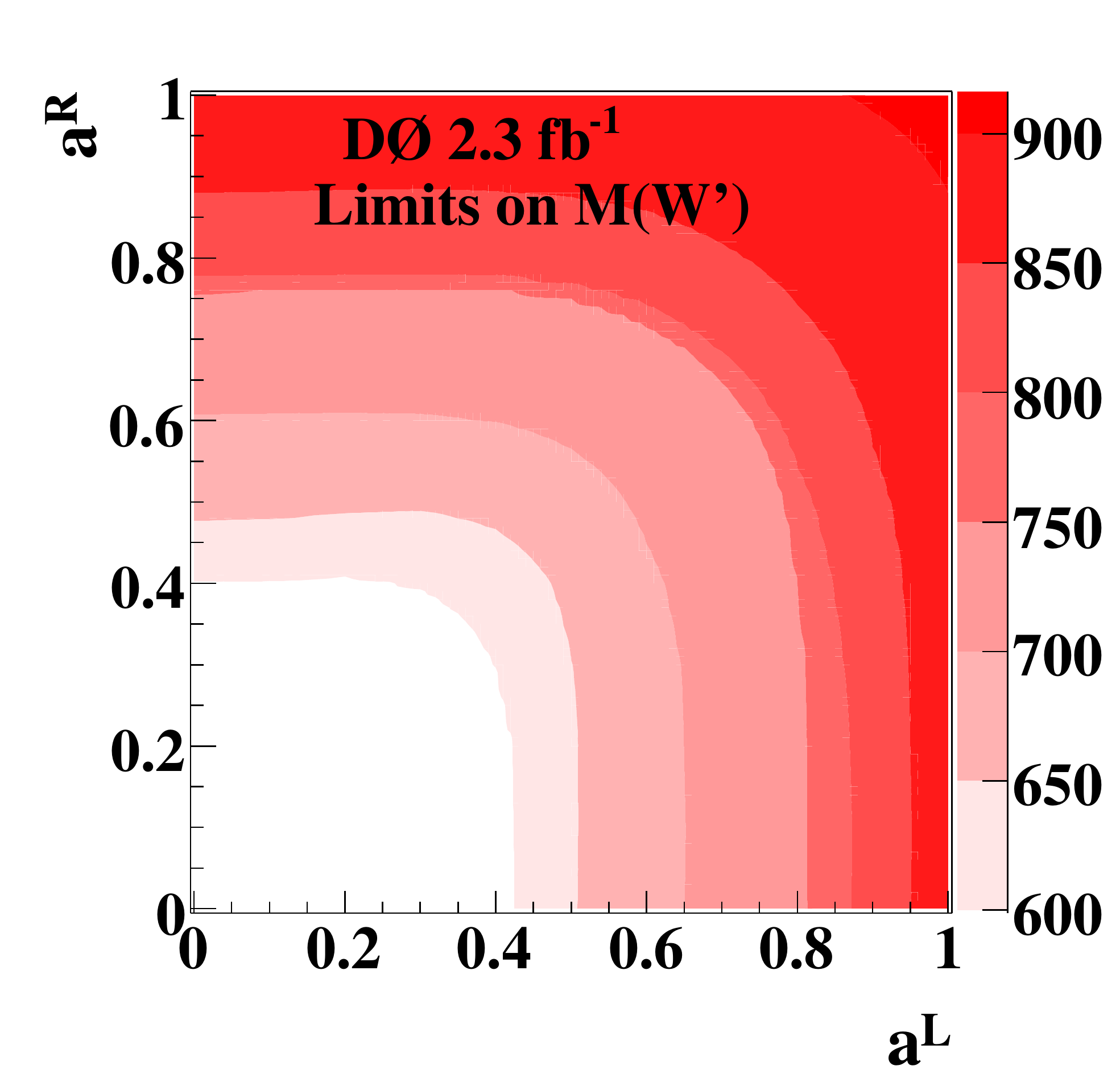}
 \caption{\label{fig:wp-lim}
   Limits on $W'$ boson mass (in GeV) as a function of its couplings.}
\end{minipage}
\end{figure}

Limits are set using a Bayesian procedure at 95\% CL for generic $W'$ mass and couplings (shown in Fig.~\ref{fig:wp-lim})
and for standard benchmark scenarios.
For left handed $W'$ boson, where the interference with the SM accounts for up to
a third of the production and is fully taken into account, D0 find $\mwp>863\GeV$.
For right handed $W'$ bosons, the limit can depend on whether the existence of a right-handed 
neutrino with a mass below the $W'$ mass. This proves to be a small effect: the
limit is $\mwp>885\GeV$ for $\mnr<\mwp$ and $\mwp>890\GeV$ otherwise.
For $a^L=a^R=1$ the limit is $\mwp>885\GeV$.

\section{Measurement of the top quark charge}

When reconstructing the \ttbar\ final state, the $b$ quark is combined
with the $W^+$ boson candidate (i.e. a combination of $qq'$ or $l\nu$)
and the \bbar\ with the $W^-$ to form top quarks
with an electric charge of $\pm\frac{2}{3}e$.
This can be seen as a working assumption, 
as in most measurements there is no attempt to distinguish between 
the $b$ jet and \bbar\ jet.
Thus an alternative was suggested in 1998~\cite{bib:chang} that 
the particle found with mass $\approx172\GeV$ has charge $\pm\frac{4}{3}e$,
so that the real top quark is heavier and the standard reconstructions are wrong.
Though this scenario has since been disfavored, we seek direct
experimental evidence that refutes it.

CDF recently reported~\cite{bib:qtop} a preliminary study of a more
generic scenario, where only a fraction of the particles of mass $\approx172\GeV$ 
(that decayed to $bW$) have a charge of $\pm\frac{4}{3}e$.
To measure the fraction of these exotic particles, $f+$, 
the $W$ boson candidates and the $b$-tagged jets are matched
using a kinematic fit, and $b$ jets are separated from \bbar\ jets 
using a jet charge.

The measurement was done in the lepton+jets channel.
To reduce the number of possible jet--parton assignments, and thus
improve the matching between $b$ jets and $W$ boson candidates,
only events with two $b$ tags are used in this study.

In the kinematic fit, the 4-vectors of the observed objects are varied
according to their experimental resolutions and under the constraints
of the known $W$ boson and top quark masses. A $\chi^2$ test statistic
is minimized over all possible jet--parton assignments.
Only events where a good assignment, with $\chi^2<9$, was found are
analyzed further.
The efficiency of this cut is 53\%, and 83\% of the selected assignments
are correct.

The jet charge is defined as
\begin{equation}
Q = \frac{\sum\limits_i \left(\vec{p}_i\cdot\vec{p}_j\right)^{0.5}Q_i}{\sum\limits_i \left(\vec{p}_i\cdot\vec{p}_j\right)^{0.5}},
\end{equation}
where $i$ runs over the tracks in the jet, with $Q_i$ the charge
and $\vec{p}_i$ the momentum 3-vector of the jet's $i$-th track, and $\vec{p}_j$ is the
jet momentum 3-vector.

The performance of the jet charge was calibrated using control data enriched in
\bbbar\ events with a tag and probe technique, and in eight bins of jet transverse energy.
The enrichment was done by $b$ tagging both jets using displaced vertices,
and by requiring a tag muon in one of the jets. The charge of the muon tags, with
an easily-modeled accuracy, the flavor of the jet that contains it.
The purity of the samples is evaluated using the \pt\ of the muon relative to the 
axis of the tag jet and using the mass of the displaced vertex in the probe jet.
The jet charge correctly identifies whether a jet originated from a $b$ or \bbar\ quark
61\% of the time. 
The distribution of the jet charge, multiplied by charge of the $W$ boson matched
to it so as to reconstruct the sign of their progenitor top quark, is shown
in Fig.~\ref{fig:jetQ}.

\begin{figure} [htbp]
\begin{minipage}[t]{0.48\linewidth}
 \centering
 \includegraphics[width=\textwidth]{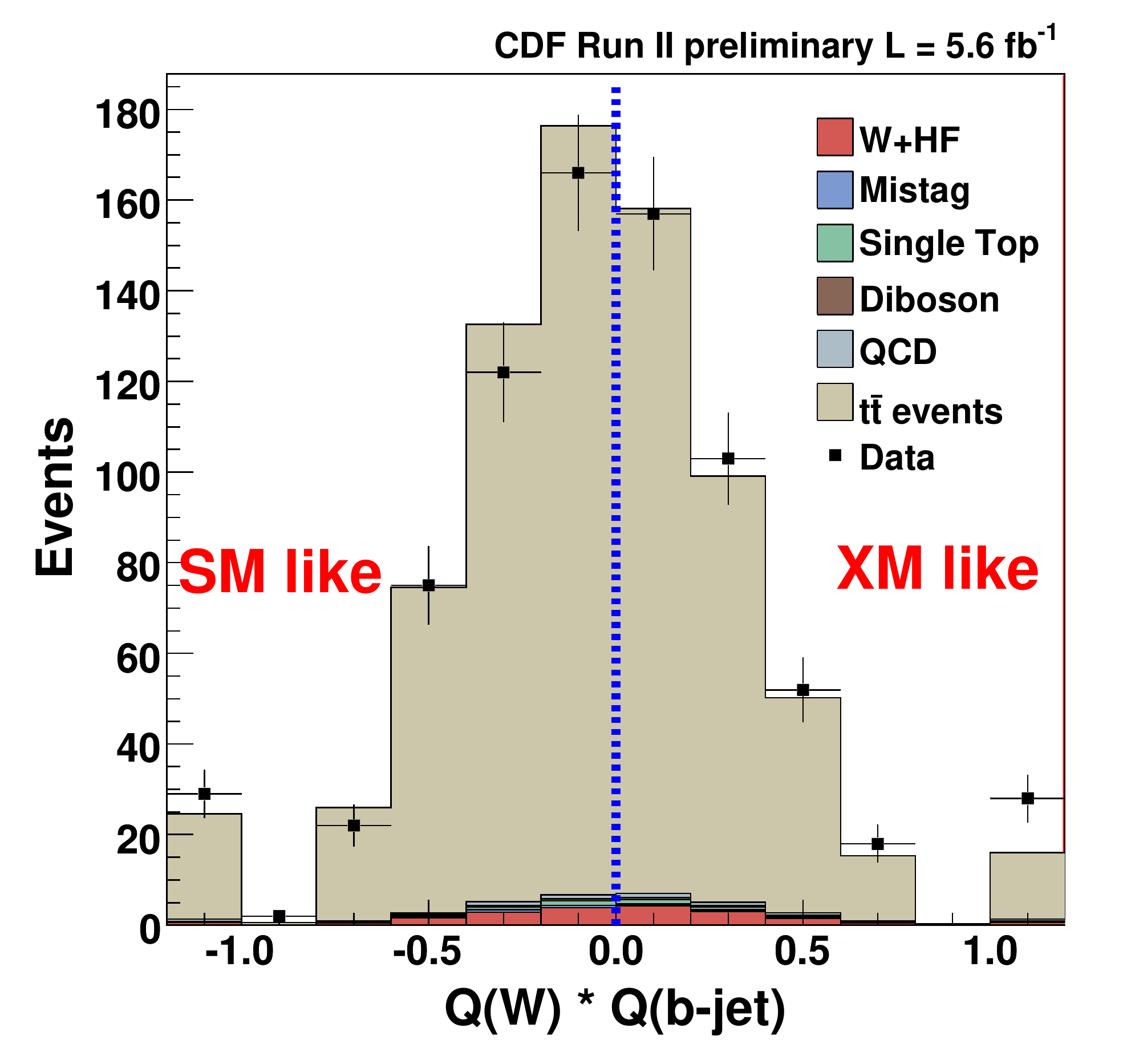}
 \caption{\label{fig:jetQ}
   Distributions of jet charge signed by $W$ boson charge and sample composition.
   ``XM'' refers to the exotic particle scenario described in the text.}
\end{minipage}
\hspace{0.03\linewidth}
\begin{minipage}[t]{0.48\linewidth}
 \centering
 \includegraphics[width=\textwidth]{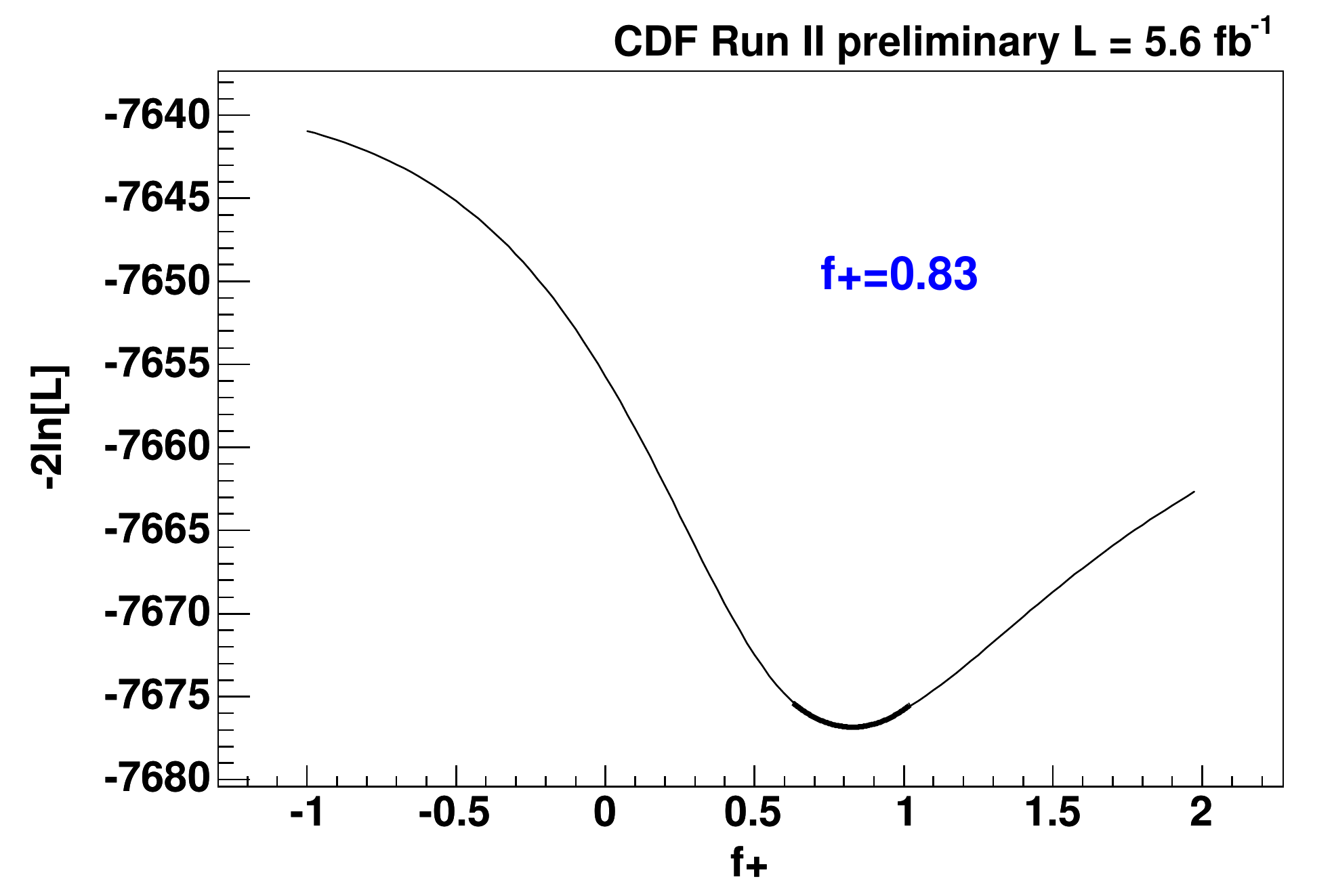}
 \caption{\label{fig:fplus}
   Profile likelihood as a function of $f+$.}
\end{minipage}
\end{figure}

The likelihood for each event is calculated, accounting for $b$-jet misidentification
and/or mistaken jet--parton assignments, as a function of $f+$. To measure $f+$
a profile likelihood test statistic is built from the per-event likelihoods
by maximizing the sample likelihood with respect to the sources of systematic
uncertainty, represented by nuisance parameters. The profile likelihood as a function
of $f+$ is shown in Fig.~\ref{fig:fplus}.

The measured value is consistent with the SM, with a $p$ value of 0.134,
but is inconsistent with the exotic charge ($\pm\frac{4}{3}e$) scenario,
with a $p$ value of $1.4\cdot10^{-4}$. Thus the alternative scenario
of Ref.~\cite{bib:chang} is ruled out at the 95\% CL.
This preference for the SM was also quantified in terms of a Bayes factor.
CDF find $2\ln{\mathrm BF}=19.6$, that is, that the data favors the SM
very strongly.

\section{\boldmath $W$ boson helicity in top quark decays}

The SM predicts $\BR\left(t\to bW^+\right)>99.8$\%, which is supported by
measurements so far~\cite{bib:rb}. This prediction can be broken down by
helicity states: the prediction 
for the fraction of left-handed $W^+$ bosons in this decay is $f_-=31.1$\%, 
for the fraction of longitudinally-polarized $W^+$s $f_0=68.7$\%, and
for the fraction of right-handed $W^+$s $f_+=0.17$\%~\cite{bib:WhelNNLO}.
The latter is suppressed due to the $V-A$ nature of charged weak
current interactions.
The experimental uncertainties are smaller than the uncertainties
on these calculation, so the measurements can not be used to constrain
the parameters of the SM. 
Instead they serve to probe for effects beyond the SM.

We distinguish between the helicity states by reconstructing \cts,
the angle between the direction from which the top quark entered the decay
and the direction of the outgoing up type fermion in the $W$ boson's rest frame.
The distributions of \cts\ for various helicity states are
shown in Fig.~\ref{fig:theta_star}.

\begin{figure} [htbp]
\begin{minipage}[t]{0.43\linewidth}
 \centering
 \includegraphics[width=\textwidth]{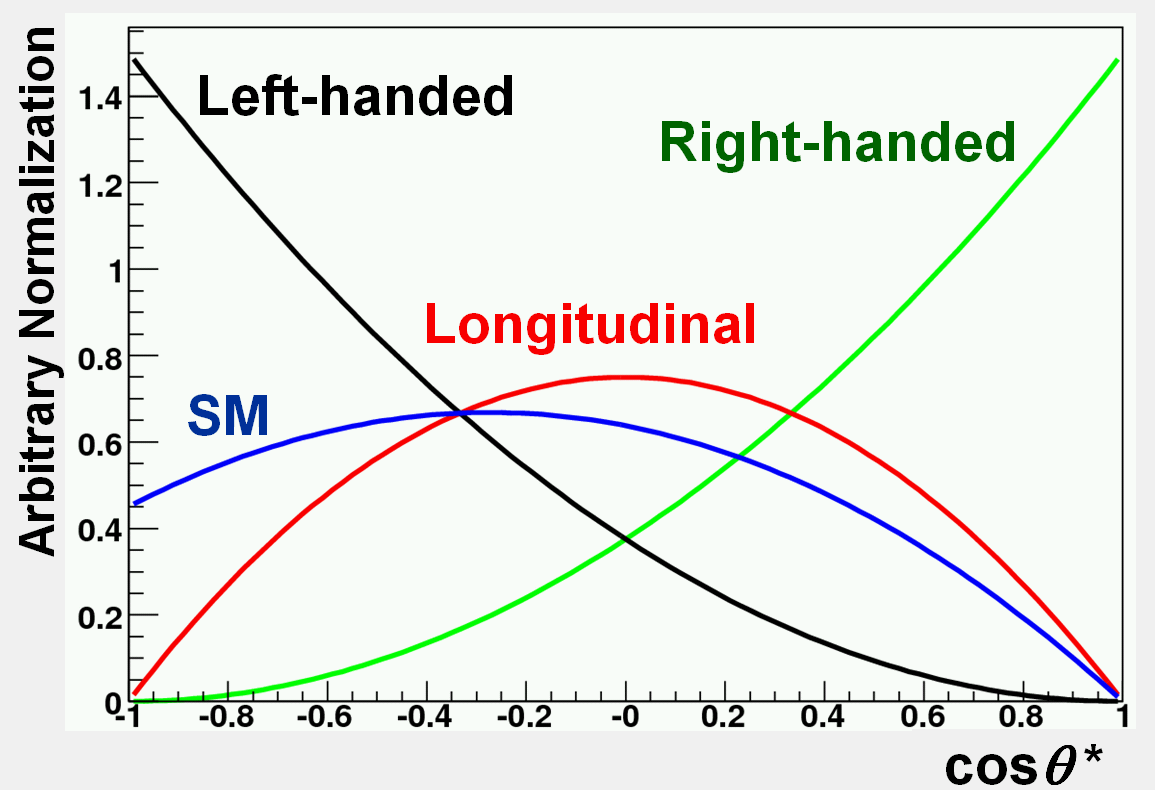}
 \caption{\label{fig:theta_star}
   Distributions of production-level \cts\ for the different helicity states and the SM.}
\end{minipage}
\hspace{0.03\linewidth}
\begin{minipage}[t]{0.53\linewidth}
 \centering
 \includegraphics*[width=\textwidth,viewport=00 0 515 290]{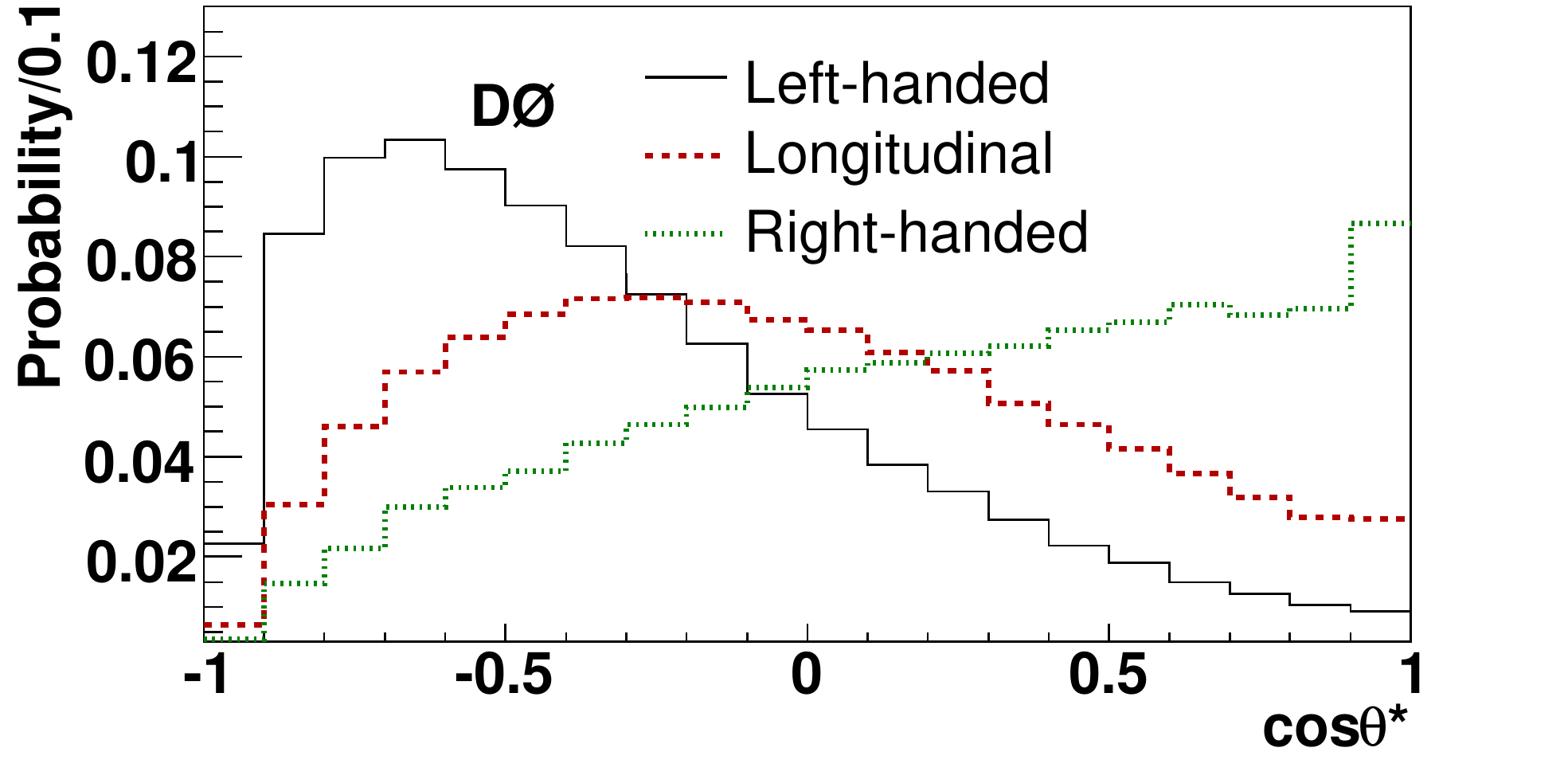}
 \caption{\label{fig:theta_ljets}
   Distributions of the reconstructed \cts\ for the different helicity states and the SM
   in the lepton+jets channel in D0.}
\end{minipage}
\end{figure}

The D0 Collaboration recently published~\cite{bib:WhelD0} a measurement of the $W$ helicity.
The measurement was performed in both the lepton+jet channel and in the dilepton channel, 
where both $W$ bosons in $\ttbar\to W^+bW^-\bar{b}$ decay leptonically. 

In the lepton+jet channel, selection and reconstruction,
through a kinematic fitter, are similar to those described above.
In particular, a discriminant is used to enrich the data in \ttbar\ events.
The excellent
reconstruction of \cts\ from the leptonically decaying top quark
can be seen by comparing Figs.~\ref{fig:theta_star} and~\ref{fig:theta_ljets}.
The deficit seen in Fig.~\ref{fig:theta_ljets} at $\cts\approx-1$ is due to
acceptance effects rather than imperfect reconstruction. In that region the 
angular separation between the lepton and the $b$ jet (from the same $t\to Wb$ decay)
is too small for them to be reconstructed as separate objects.
For the hadronically decaying top quark, only $\left|\cts\right|$ can be reconstructed,
since the flavor of the jets is not readily available. Nevertheless, the information
from the hadronically decaying top quarks is useful to constrain $f_0$.

In the dilepton channel, events are selected if they contain two
isolated leptons (again this refers to electrons and muons, possibly through
an intermediate $\tau$ lepton) of opposite charge and at least two jets.
Discriminants are built for each channel ($ee$, $e\mu$ and $\mu\mu$) 
and used to enrich the data in \ttbar\ events.
Due to the two unmeasured neutrinos, the kinematics of the system are under constrained.
A probabilistic reconstruction is used: the objects are varied within their
experimental resolutions and for each variation the mass constraints are used
to find the possible neutrino momenta. Up to eight solutions are possible.
The two \cts\ values are calculated for each solution in 500 variations of the objects,
and their averages are the reconstructed \cts\ values.
The resulting separation is somewhat weaker than in the lepton+jets channel.

The CDF Collaboration has recently measured the $W$ helicity in the dilepton
channel using a similar technique~\cite{bib:WhelCDF}. Notable differences are:
(a) instead of using custom discriminants, CDF enrich the data in \ttbar\ using
$b$ tagging, and (b) CDF use additional criteria to select the most
likely solution out of the up to eight solution for each variation.

In both analyses, the $W$ helicity fractions are extracted by fitting
a linear sum of the templates for each helicity state (e.g. those shown 
in Fig.~\ref{fig:theta_ljets}) and for background to data. 
The results of the fits are shown in Fig.~\ref{fig:WhelRes}.
CDF then further calibrates their method, finding corrections of $1$--$2$\%
which yield $f_0=0.78\pm0.20$ and $f_+=-0.12\pm0.10$.
D0 find $f_0=0.67\pm0.10$ and $f_+=0.02\pm0.05$.

\begin{figure} [htbp]
  \centering
  \includegraphics*[width=0.41\textwidth,viewport=10 23 520 490]{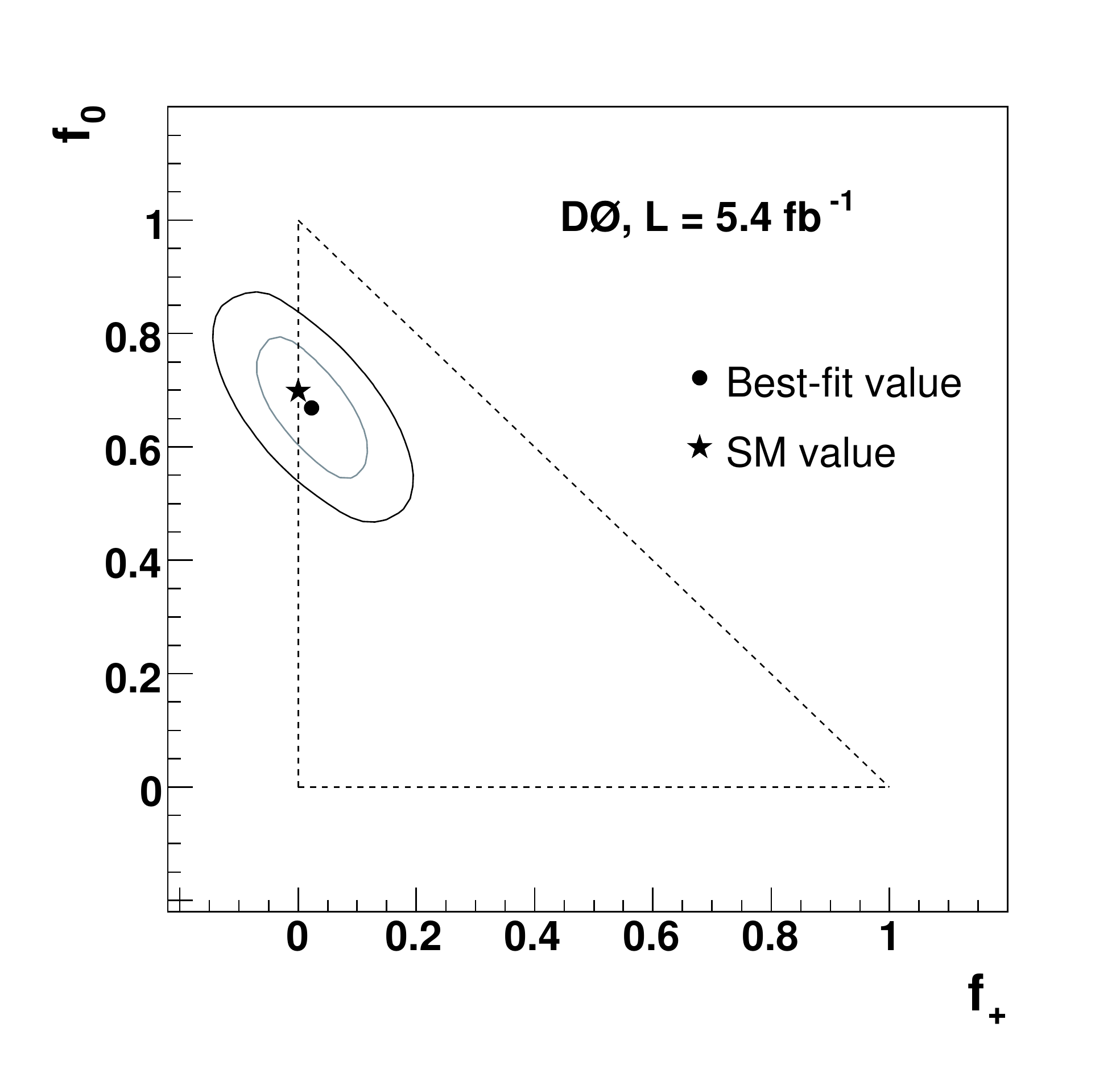}
  \hspace{0.02\linewidth}
  \includegraphics[width=0.55\textwidth]{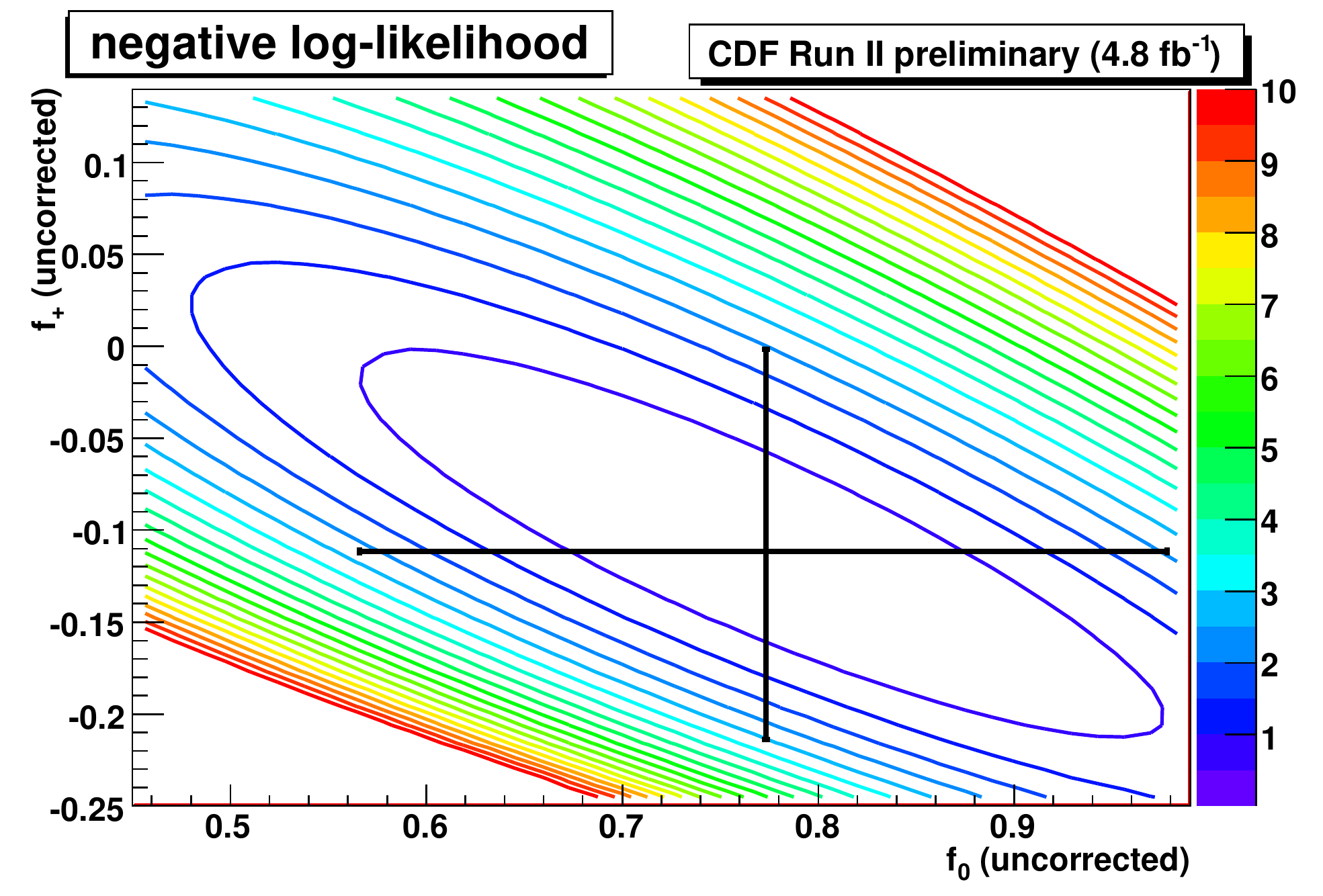}
 \caption{\label{fig:WhelRes}
   Fitted $W$ boson helicities in top decays from D0 (left) and CDF (right).
   Note the different axis definitions. In the left plot, the dashed triangle
   shows the physically allowable region, while the blue and black curves
   show the $1\sigma$ and $2\sigma$ contours. In the right plot, the black lines
   represent the statistical uncertainties, and the colored lines show contours.}
\end{figure}

\section{\boldmath Direct measurement of the $t$--$\bar{t}$ mass difference}

From the CPT theorem we learn that for any canonical~\footnote{local, Lorentz-covariant, 
and with a Hamiltonian that is bounded from below} quantum field theory, 
the mass of particles and their antiparticle are identical.
Due to QCD color confinement, quark masses are not directly accessible.
The only exception is the top quark, which decays before it is confined.
Both collaborations measure the mass difference between top and antitop 
quarks~\cite{bib:deltam}, as a probe for BSM effects.

The mass difference is measured in the lepton+jets channel, with
at least one $b$-tagged jet. The charge of the lepton separates
the top from the antitop quark. 
An interesting experimental aspect is that the calorimetry's
response to $b$ and \bbar\ jets may differ.

The D0 measurement was a variation of their mass measurement using
the matrix element technique~\cite{bib:dzme}, with the signal
generated a modified version of \pythia. 
They find $m_t - m_\tbar = 3.8\pm3.7\GeV$, as shown
in Fig.~\ref{fig:dzdeltam} for each lepton flavor.

The more recent CDF measurement uses a much larger data sample.
Signal is generated using \madgraph~\cite{bib:madgraph}
and the events are reconstructed using a kinematical fitter.
Unlike the analyses described above, in this analysis for each
event the two best solution are retained (see Fig.~\ref{fig:cdfdeltam}).
The mass difference is measured with an unbinned likelihood fit
with the data split into channels by the number of $b$ tags.
CDF find $m_t - m_\tbar = -3.3\pm1.7\GeV$.

\begin{figure} [htbp]
\begin{minipage}[t]{0.53\linewidth}
 \vspace{4pt} 
 \centering
 \includegraphics[width=0.49\textwidth]{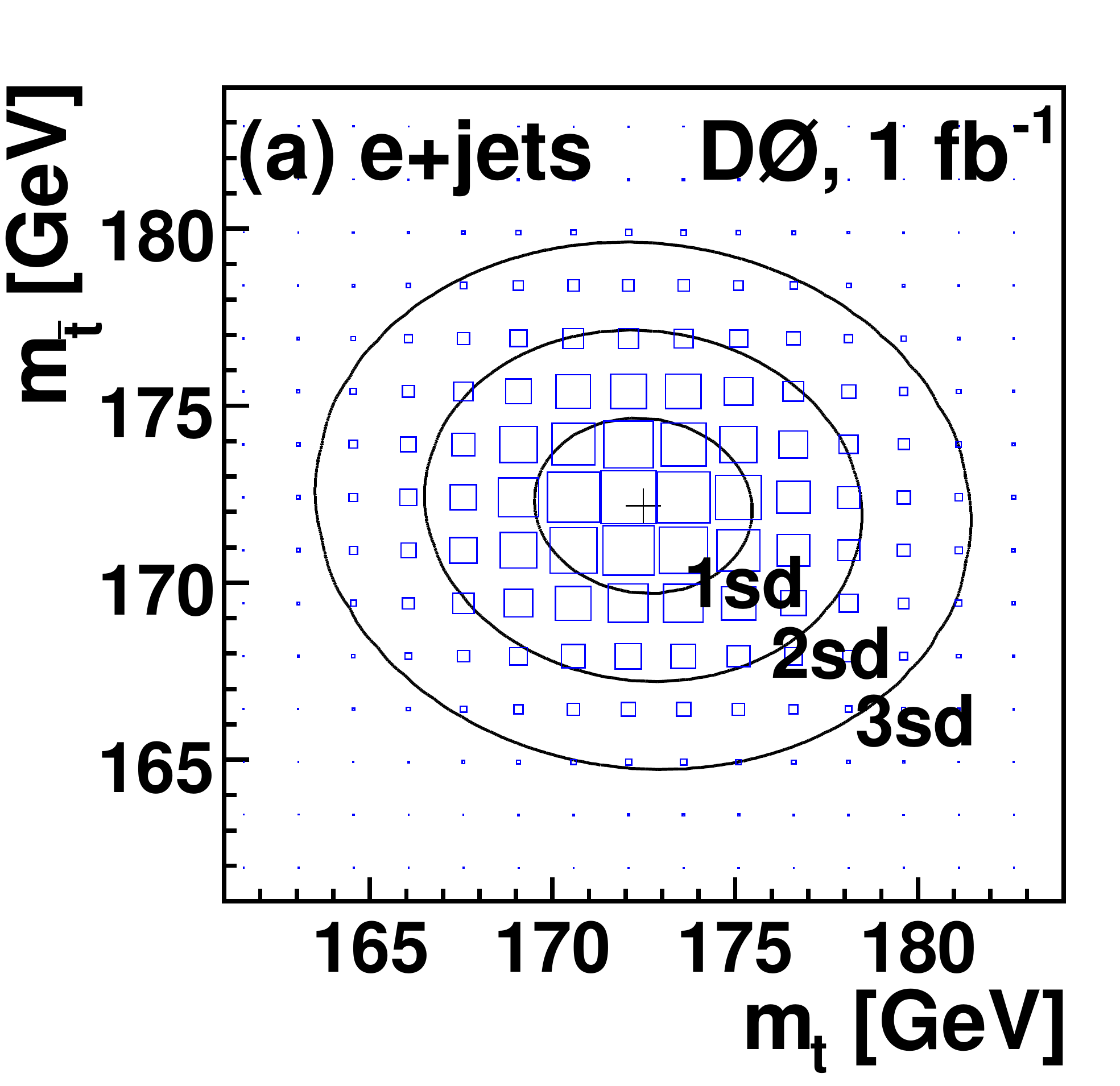}
 \includegraphics[width=0.49\textwidth]{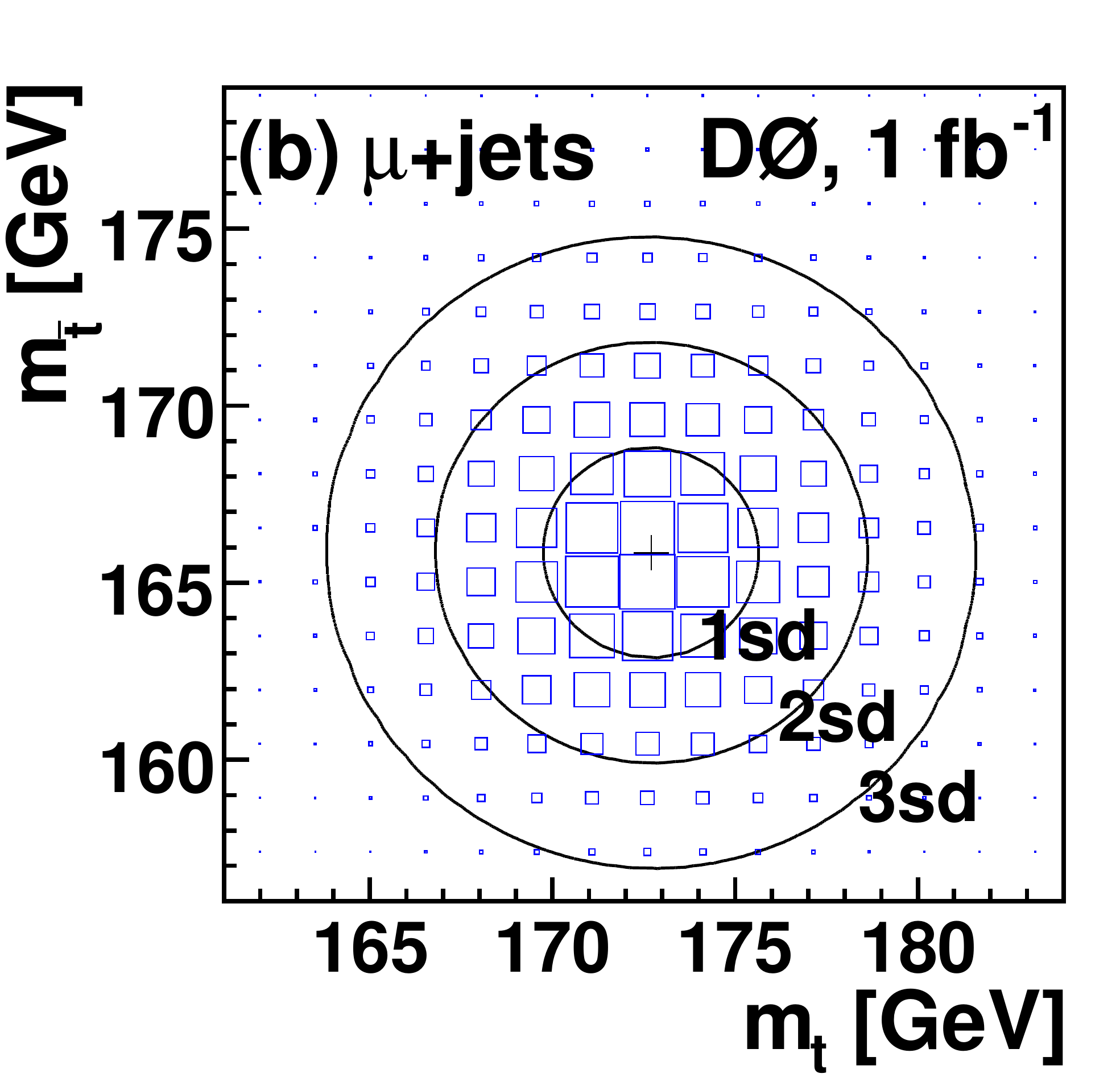}
 \caption{\label{fig:dzdeltam}
   Likelihood of $m_t, m_\tbar$ from D0. The black curves show contours of equal probability.}
\end{minipage}
\hspace{0.03\linewidth}
\begin{minipage}[t]{0.43\linewidth}
  \vspace{0in} 
 \centering
 \includegraphics[width=\textwidth]{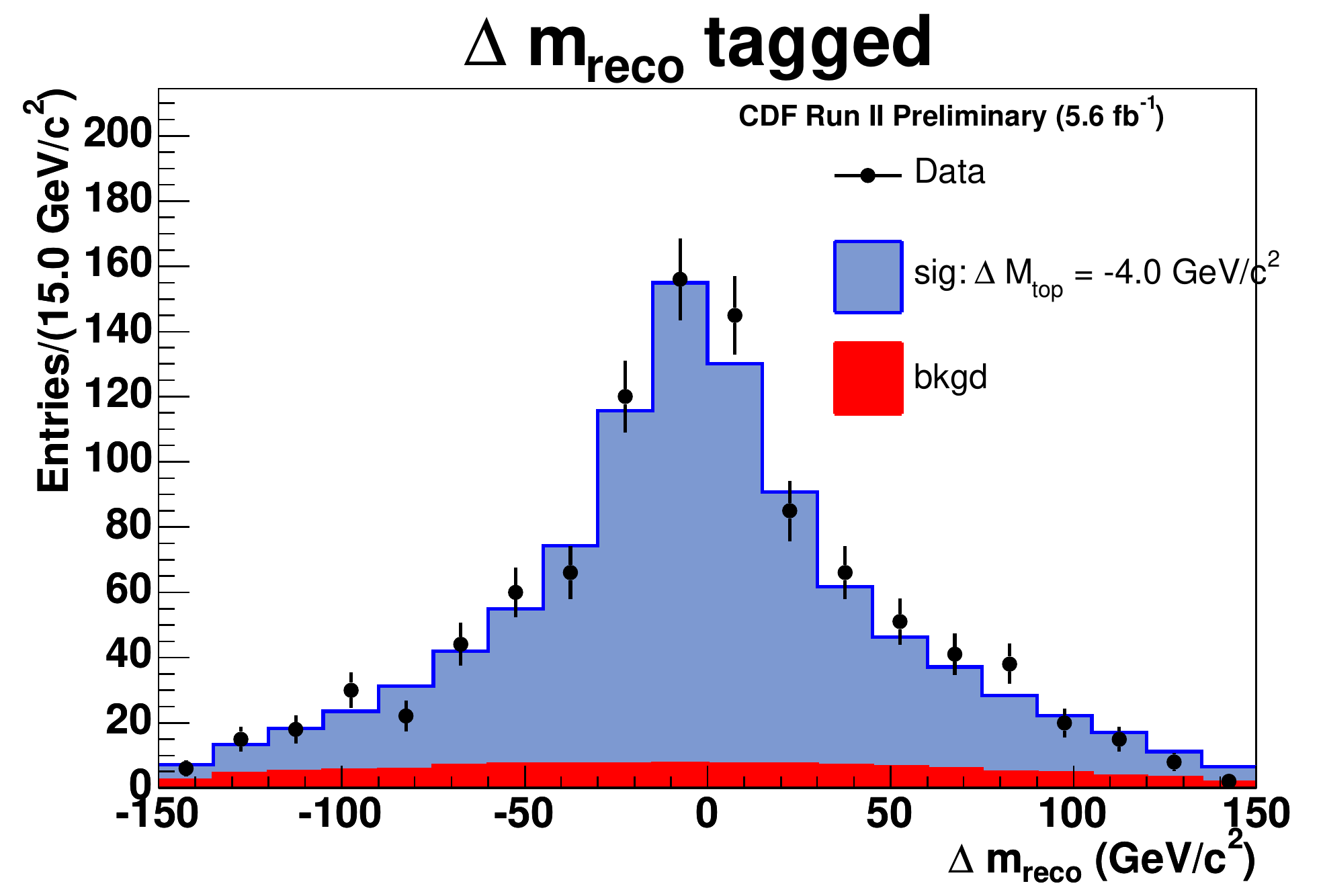}
 \caption{\label{fig:cdfdeltam}
   Reconstructed mass difference from CDF.}
\end{minipage}
\end{figure}


\section{Conclusion and Summary}

The standard model describes Tevatron top data well, with 
a caveat presented in Ref.~\cite{bib:schwarz}.
The Tevatron experiments search for a wide variety of 
new physics in top events, using both direct and indirect
searches. The latest examples were reported here.

\end{document}